# Electronic transport properties of indolyl spirooxazine/merooxazine-based light-driven molecular switch: The effect of amino/nitro substituents

H. Zhao[a, *], Y.Q. Xu[a], W.K. Zhao[a], K. Gao[a]

[a] *School of Physics, state Key Laboratory of Crystal Materials, Shandong University, Jinan 250100, China*

[*] Corresponding author at: School of Physics, state Key Laboratory of Crystal Materials, Shandong University, Jinan 250100, China. Tel.:+8618769785760. *E-mail* address: sduwuli2010zh@126.com (H. Zhao).

## Abstract

By applying non-equilibrium Green's function formulation combined with first-principles density functional theory, we explore the electronic transport properties of indolinospironaphthoxazine (SO)/indolinomeronaphthoxazine (MO). The results indicate that the MO allows a far larger current than the SO. The substituent group can cause the shift of the energy levels. Higher ON/OFF current ratio can be obtained if either amino or nitro substituent is placed at the position of naphthalene moiety. Our results suggest that such molecular wires can generally display perfect switching function and the efficiency can be tuned flexibly by adding certain substituent groups to the molecules.

**Keywords:** Molecular electronics; Density functional theory; Optical molecular switch; Nonequilibrium Green's function; spirooxazine

## 1. Introduction

It is clearly known that today's semiconductor-based microelectronics fabrication has a large scale of physical limitations in sight, but molecular electronic devices are considered as appropriate candidates for nanometer electronics [1]. Lots of novel physical properties like negative differential resistance (NDR), electronic switching in molecular devices, memory effects and molecular rectification have been found in various systems including organics, carbon nanotubes, and DNA, etc [1, 2-8]. The most prominent among these is molecular switch, which has highly potential for applications to optically rewritable data storage [7], logic devices and so on. Molecular switching devices can be mainly divided into two categories. One is the electronic switching devices which control the conversion between high conductance (ON) and low conductance (OFF) states by external triggers such as the electric field [2], the tip of scanning tunneling microscopy (STM) [5], and the redox process [6], etc. However, these triggering means are not ideal since they may interfere greatly with the function of a nano size circuit and limit the real applications. The other is optical switching devices which exist in two thermally sufficiently stable states with ON state and OFF state by light [4]. Light is a very attractive external stimulus for such switches including short response time, the ease of addressability and compatibility with a wide range of condensed phases [9]. In this regard, albeit there have been many theoretical as well as experimental investigations pursued on optical switches, the number of molecular systems having the switching feature are still limited.

Derivatives of spiropyran and spirooxazine, which are characterized by marked photochromism [8, 10, 11] have been two important classes of photochromic compounds recently. The only difference in structure between these two compounds is that the pyran's C=CH is substituted by C=N [12]. Although the spiropyran family has been more profusely studied, spirooxazines are more promising from the commercial point of view because their photoresistance is greater due to the stabilization induced by the nitrogen atom of the oxazine [13, 14]. However, spirooxazines derivatives' photochromic property was firstly reported by Fox [15], who synthesized the first spirooxazine compound 1,3,3-trimethylspiro [indoline-2,3'-[3H]naphtha[2,1-b][1,4]oxazine]. And there have been lots of theoretical and experimental researches on the photochemical and photophysical behaviors of SO, but we focus on the electronic transport properties of this molecule. Fig. 1 shows that the colorless SO, upon irradiation with ultraviolet (UV) light, isomerizes to the purple MO through heterolytic cleavage of the spiro carbon–oxygen bond. The process is thermally and photochemically reversible. This may provide a potential molecular switch in future nanoelectronics applications.

Besides, in recent years, carbon nanotubes (CNTs) have attracted great attention [16-18]. They are considered as one of the most promising low-dimensional materials for making molecular devices for their high chemical stabilities and remarkable electronic [19] and thermal [20] properties. Recent experiments suggest that carbon-based electrodes are a viable alternative to the more commonly used noble metals [21, 22]. In the present work, by applying nonequilibrium Green's function (NEGF) combined with density functional theory (DFT), we investigate the electronic transport properties of three different groups of spirooxazine derivatives' optical molecular switches with SWCNT electrodes. And we get that the amino-substituted group has the highest ON/OFF ratio.

## 2. Model and method

The geometrical optimizations, current–voltage (*I-V*) characteristics and the electronic transport properties calculations have been performed by using the first principles packages Atomistix ToolKit (ATK) [23, 24], which is based on the NEGF and DFT techniques. This package, as implemented in the well tested TranSIESTA-C method, is capable of fully self-consistently modeling the electrical properties of nanoscale device that consists of an atomic scale system coupling with two semi-infinite electrodes. The SO and MO are sandwiched between two (5, 5) armchair carbon nanotube electrodes. In NEGF theory, such a two-probe system is divided into three regions: left electrode, central scattering region, and right electrode. The semi-infinite electrodes are calculated separately to obtain the bulk self-energy. The scattering region includes the SO (MO) and five surface-atomic layers of the carbon atoms. The amount of layers is large enough to screen the perturbation effect from the central scattering region for each electrode [25]. The distance between the two electrodes is 14.40 Å, which is chosen to fit best with the length of the optimized closed-form molecule. The open form has more degrees of freedom and can fit well in longer or shorter distances between the electrodes. From the model in the Fig. 1 (c) and (d), we can find that the central molecule is linked with the SWCNTs directly. As shown in the Fig. 1(a) and 1(b), such converse reaction can take place between (a) and (b) upon photoexcitation. When SO is irradiated with ultraviolet ($UV_1$), an excitation state is achieved and SO inverts into MO. The irradiation with another frequency $UV_2$ or heat turns MO into SO, and the reversible reaction by UV and heat creates a continuous inversion movement, as shown in Fig. 1.

The geometry is optimized by minimizing the atomic forces to be smaller than 0.05 eV/Å. The remaining part of SWCNT is fixed as C-C bond length of 1.422 Å. The geometry optimization and electronic transport properties are implemented in the ATK software package (2008.10.0). In the calculations, the exchange-correlation potential is described by the generalized gradient approximation (GGA) in Perdew-Burke-Ernzerhof (PBE) method [26]. The core electrons are modeled with the Troullier-Martins nonlocal pseudo potential [27], and the valence electrons are expanded in a single-$\zeta$ plus polarization basis set (SZP) for all atomic species to achieve a balance between calculation efficiency and accuracy [28]. The resolution of the real space grid is determined by an equivalent plane which is set to be cutoff of 150 Ry and the Brillouin zone is sampled by a 1×1×300 points within the Monkhorst-Pack $k$-points sampling scheme. In addition, to avoid the interaction between the molecule and its periodic images, a large super cell dimension (20 Å) in the plane perpendicular to the transport direction is used.

The transmission function $T(E,V)$ of the system is the sum of transmission probabilities of all channels available at energy $E$ under external bias $V$ [29]:

$$T(E,V) = Tr[\Gamma_L(V) G^R(E,V) \Gamma_R(V) G^A(E,V)],$$

where $G^{R,A}$ is the retarded and advanced Green's functions, and coupling functions $\Gamma_{L,R}$ are the imaginary parts of the left and right self-energies, respectively. The zero-bias conductance can be obtained from the transmission probability at zero bias [29]:

$$G = G_0 T(E, V=0),$$

where $G_0 = 2e^2/h$ is the quantum unit of conductance, $h$ is Planck's constant, and $e$ is the electron charge. The current ($I$)-voltage ($V$) characteristics are obtained from the Landauer–Bütiker formulation [29]:

$$I(V) = \frac{2e}{h} \int [f(E - \mu_L) - f(E - \mu_R)] T(E,V) \, dE,$$

where $f$ is the Fermi function, and $\mu_{L,R}$ are the electrochemical potentials of the left and right electrodes, $\mu_{R/L}(V) = E_F \pm eV/2$. $E_F$, the Fermi energy, is set to be zero in our calculation.

Contributing to the total current integral, the energy region [ $\mu_L(V)$, $\mu_R(V)$ ] is called the bias window.

## 3. Results and discussion

Three different SO/MO groups are considered in the report. SO and MO are our main focus, and the effect of two substituents is also investigated.

### 3.1 Transport properties of the SO and MO

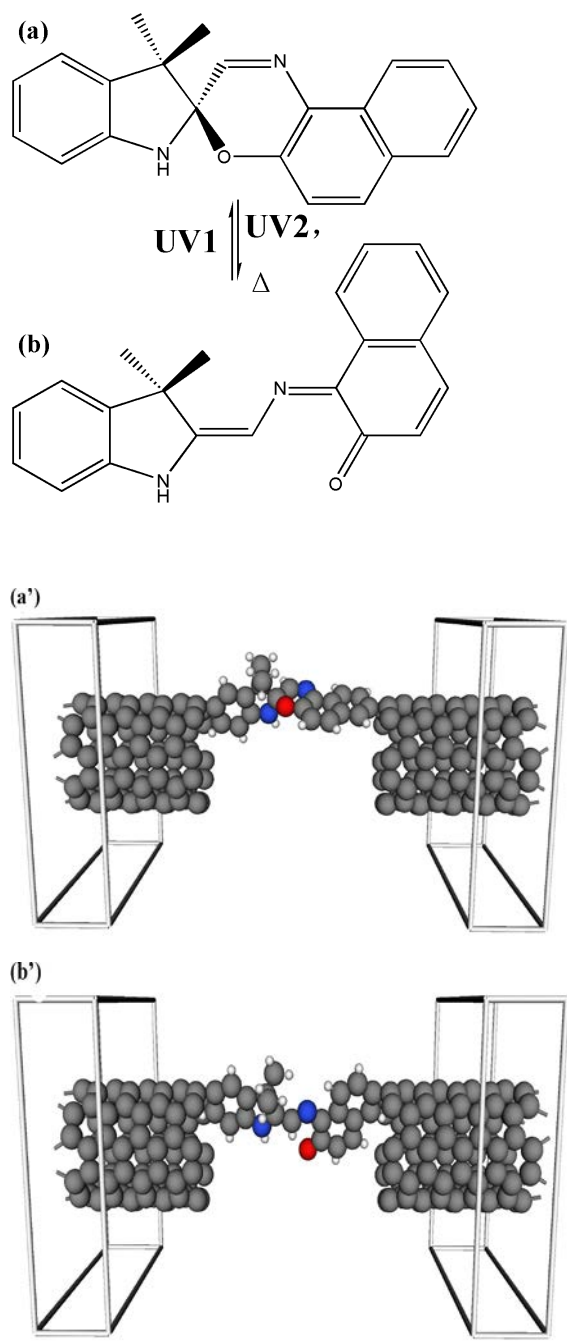

**Fig. 1.** Conversion between (a) SO and (b) MO. Schematic descriptions of the (a') SO and (b') MO devices. The region in the box indicates the electrodes, and the one between two boxes is the scattering region. The gray, red, white, and blue spheres represent C, O, H, and N atoms, respectively.

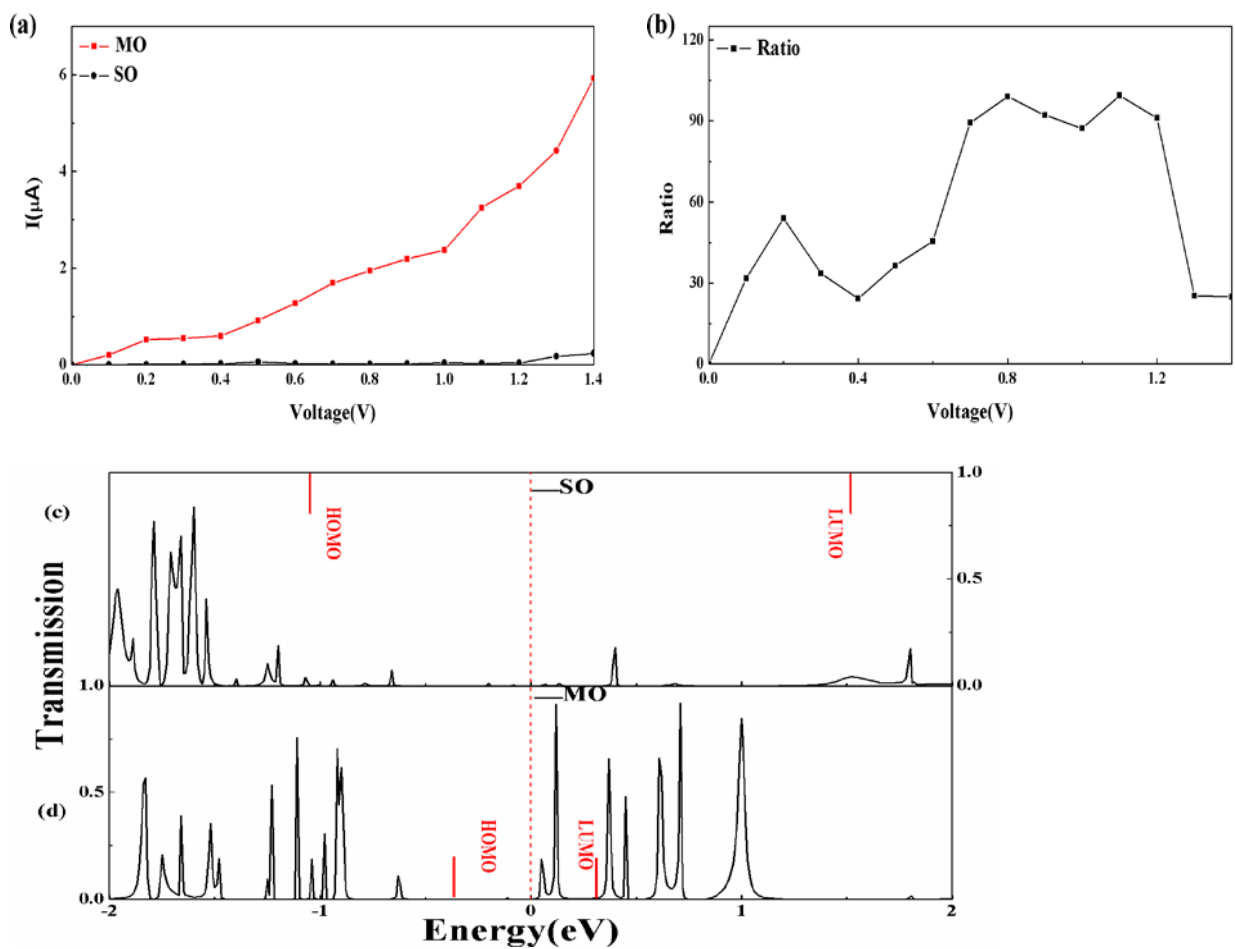

**Fig. 2.** (a) Calculated current as a function of the applied bias for SO and MO. (b) The ON/OFF ($I$ (MO)/$I$ (SO)) ratio. Transmission spectra, HOMO and LUMO energy levels for (c) SO and (d) MO at zero bias voltage. The $E_F$ is set to be zero.

From the Fig. 2(a), we can discover that the $I-V$ curves of SO and MO are really different when the bias is in the range [0 V, 1.4 V]. The current across MO is really larger than that crosses SO. Current of MO increases quickly, on the contrary, SO's current increased much slower until the bias is set to be 1.2 V. In particular, Fig. 2(b) shows that in the bias range [0.9 V, 1.2 V], the ON/OFF ratio is larger than 80 and reaches 99 at 0.8 V. Therefore, SO/MO system can be a candidate for molecular switching devices.

Molecular conductance generally follows the characteristics of molecular orbitals, particularly the frontier molecular orbitals (highest occupied molecular orbital (HOMO), lowest unoccupied molecular orbital (LUMO)), and the HOMO-LUMO gap. To understand the switching mechanism, we calculated the transmission spectrum, HOMO and LUMO energy levels at zero bias for MO and SO, respectively. Surprisingly, the transmission probability of SO is very low in the whole energy range considered, even at the energy positions of the HOMO ($E$ = -1.0515 eV) and LUMO ($E$ = 1.5217 eV), as shown in Fig. 2(c). This behavior arises since the electron can't tunnel through the junction easily because of the orthogonality of the $\pi$ systems in SO. However, the much smaller HOMO ($E$ = -0.371 eV)–LUMO ($E$ = 0.3092 eV) gap, the larger transmission probability and two transmission peaks near $E_F$ which originates from LUMO, as shown in Fig. 2(d), indicate that MO conducts electron well and can easily keep the circuit in the ON state. The reason for that characteristics of MO is that the photoreaction leads to the spiro C-O bond breaking to form C=C bond. Furthermore, indolyl and meronaphoxazine $\pi$ states are not perpendicular to each other.

Molecular projected self-consistent Hamilton (MPSH)[29] for HOMO and LUMO energy levels at zero bias voltage for MO and SO are given in Fig. 3, which illustrate intuitively the transport behaviors for the two structures. From Fig. 3(a) and 3(b), for both HOMO and LUMO of SO, the states localize on naphthoxazine and there is very small amplitude broadened on indolyl. While Fig. 3(c) and 3(d) show that HOMO and LUMO of MO distribute widely on the entire molecule. Delocalized $\pi$ molecular orbitals on the whole MO molecule significantly contribute to the electron tunneling. While in the case of SO, the extension of the $\pi$ channels is broken after the MO is lighted by ultraviolet or heat which transforms the C=C bond into the spiro C-C bond.

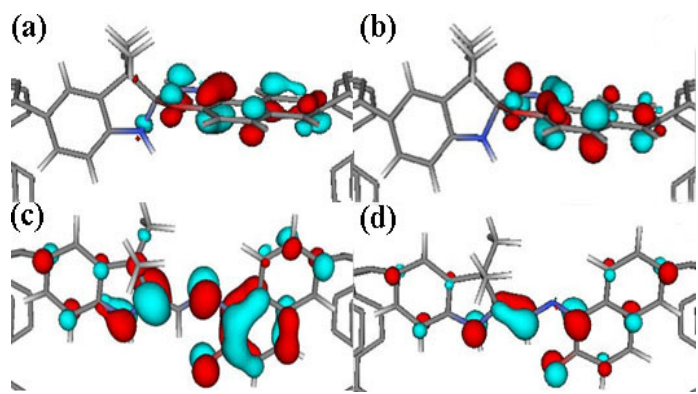

**Fig. 3.** The spatial distribution of the MPSH eigenstates of the (a) SO (HOMO) (b) SO (LUMO) and (c) MO (HOMO) (d) MO (LUMO).

When the π-conjugated indolyl and naphthoxazine are in the same plane, the contributing π orbitals will have the maximum overlap to each other and can cause the lowest resistance in the junction. For the wedge shaped SO, the spiro C-C bond between indolyl and oxazine rings leads to localization of the frontier orbitals on the naphthalene and thus weaken electron tunneling probability.

## 3.2 Influence of substituent groups

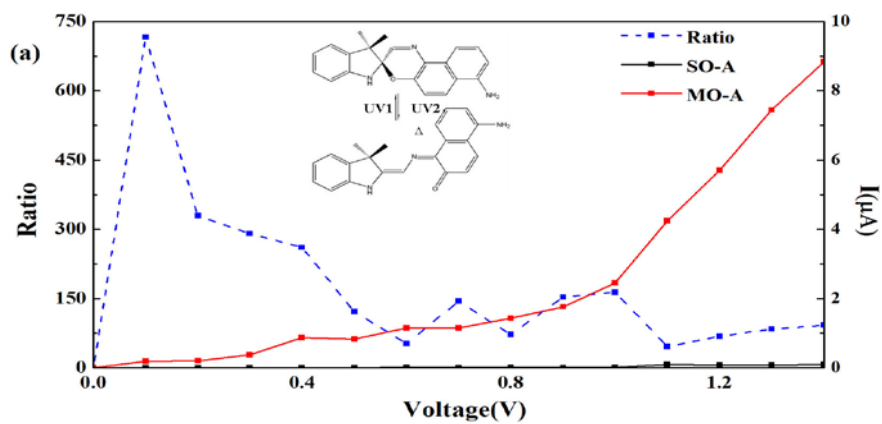

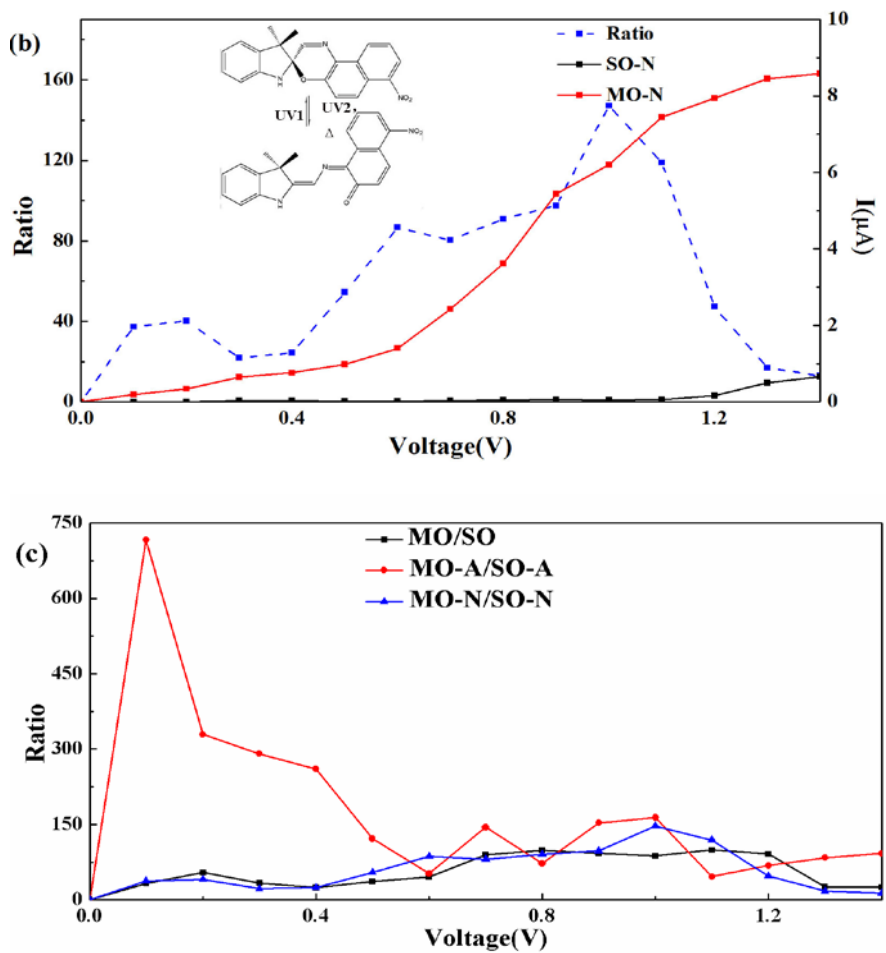

**Fig. 4.** The $I-V$ curves and ON/OFF ratios of the current for (a) MO-A/SO-A and (b) MO-N/SO-N molecule junctions. (C) ON/OFF ratios for the three systems.

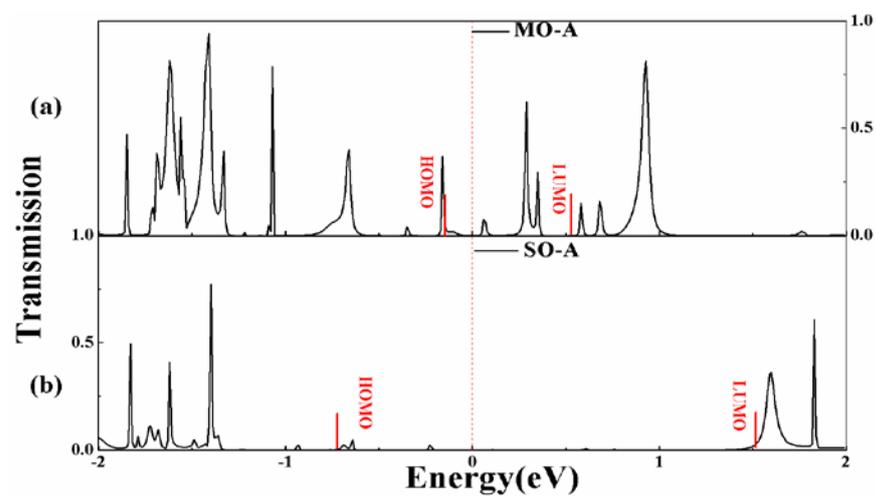

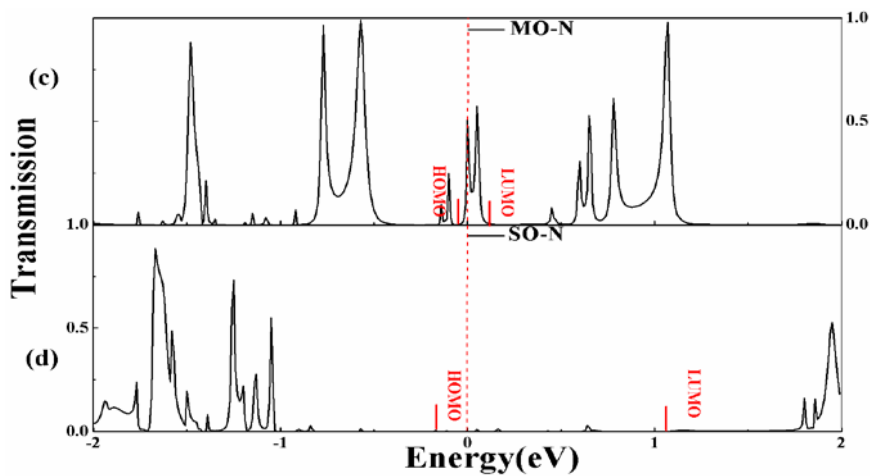

**Fig. 5.** The transmission spectra and projected HOMO and LUMO levels for (a) MO-A, (b) SO-A, (c) MO-N, (d) SO-N at the bias of zero voltage. The $E_F$ is set to be zero.

Since the electronic properties of molecular junction can be modified strongly by adding functional group such as amino, nitro, etc. to the molecules, some significant behaviors such as NDR or switching can be obtained and utilized [30, 31]. Hence, we explore the electronic transport properties of amino-substituted group SO-A/MO-A and nitro-substituted group SO-N/MO-N.

The photoreaction between SO-A and MO-A is shown in the Fig. 4(a) and the SO-N's reaction is shown in the (b) part. As the Fig. 4(a) and (b) show, the $I-V$ curves of SO-A (SO-N) and MO-A (MO-N) have similar features with SO/MO group. Therefore, ones see that the switching behavior can again be achieved by the photoreaction. And from the Fig. 4(c), it can be clearly seen that the ratio of MO-A/SO-A is much higher than the other two groups especially under the bias of 0.6 V, the ratio even reaches 716 at the bias of 0.1 V. The SO-A having only $2.314 \times 10^{-4}$ μA to tunnel through the molecular junction can explain that high ratio at the bias of 0.1 V to some degree. The corresponding results of MO-N/SO-N are given in Fig. 4(b), which also shows that the ON/OFF ratio (near 150) is higher than that of the MO/SO.

Generally, amino (NH$_2$) and nitro (NO$_2$) groups have opposite electron withdrawing capability: the former tends to lose electrons and the latter to gain [32]. Why are the switching ratios induced by the two groups not opposite but both increased? To understand the electronic transport behaviors of the two kinds of molecular junctions, we plot and analyze their transmission spectra and projected HOMO and LUMO levels in Fig. 5, respectively. In comparison with those of SO (MO), the transmission spectrum of SO-A (MO-A) (Fig. 5(a) and (b)) displays similar trends, but with larger magnitude for MO-A than MO near $E_F$. It can be induced why the current of SO-A junction is under 0.1 μA from the Fig. 5(b), where there is almost no transmission spectra from -0.6 eV to 1.4 eV. The amino substitution makes the HOMO of both SO-A ($E$ = -0.6955 eV) and MO-A ($E$ = -0.1521 eV) closer to $E_F$ and transport probability higher.

Adding a nitro group also gives rise to a highest transmission in MO-N junction (Fig. 5(c)) near $E_F$, compared to MO and MO-A cases. The LUMO, which is slightly nearer to the $E_F$ than the HOMO, goes nearer to the $E_F$ than MO. And the gap of HOMO-LUMO is much smaller than that of the MO and there are several peaks near $E_F$, which can explain why the nitro group has larger current. From the Fig. 5(a) and (b), we can see that the HOMO level in MO-A and SO-A shifts nearer to $E_F$, which can be rationalized well by the electron-loss behavior of $NH_2$ group. On the other hand, according to the Fig. 5(c) and (d), the electron-gain behavior of $NO_2$ group can explain why the LUMO level in SO-N and MO-N shifts nearer to $E_F$ well.

The amino and nitro groups' contribution to the electronic transport property can also be seen from the Fig. 6. Fig. 6(a), (b), (c) and (d) show that HOMO and LUMO of SO-A and SO-N have similar distribution with that of SO. But in the case of SO-A, there is only HOMO distribution on $NH_2$, while LUMO distribution on $NH_2$ can't be seen. As far as SO-N is concerned, only LUMO distributes on $NO_2$ substitution. So the amino and nitro substitutions have different impact on the different electronic orbitals. And from the Fig. 6(a'), (b'), (c') and (d'), we find that HOMO and LUMO of MO-A and MO-N have similar distribution with that of MO: HOMO and LUMO distribute widely on the entire molecule. The substituents causes the misalignment of the energy levels in SO-A and SO-N. The transmissions of SO-N and SO-A increase and decrease due to that, respectively, at zero bias, compared to SO. ( The trend can be seen when the Fig. 2(c), 5(a) and 5(c) are enlarged.) Additionally, compared to MO/SO group, the ON/OFF ratio increases slightly for SO-N/MO-N group because the transmission of MO-N increases a lot. For SO-A, Fig. 3(a) and Fig. 6 show that there is the highest localization of HOMO and LUMO level states on SO-A junction, which can explain why the ON/OFF ratio of MO-A/SO-A can reach the highest ratio among the groups we considered. Thus both amino and nitro groups can tune the switching effect of SO-X/MO-X group. The amino group can provide extremely high efficiency.

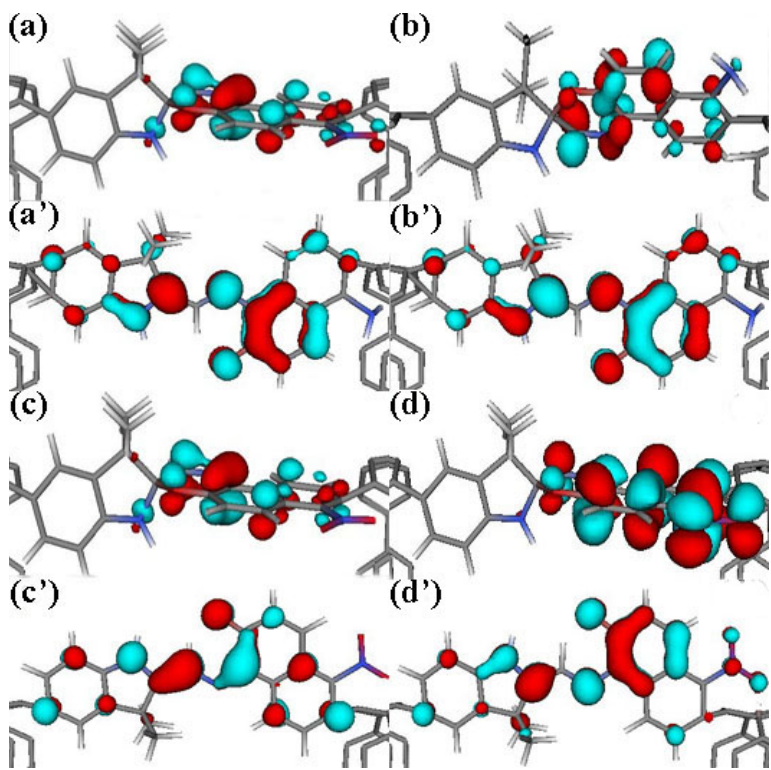

**Fig. 6.** The spatial distribution of the MPSH eigenstates of the (a) SO-A(HOMO), (a') MO-A(HOMO), (b) SO-A(LUMO), (b') MO-A(LUMO), (c) SO-N(HOMO), (c') MO-N(HOMO), (d) MO-N(HOMO) and (d') MO-N(LUMO) at the bias of zero voltage.

## 4. Conclusion

We have investigated the electronic transport behaviors of different SO/MO derivatives by applying NEGF+DFT. The $I-V$ curves indicate that all three groups can achieve a switching effect by lighting with ultraviolet. The reason is that the delocalization of frontier molecular orbitals in real space in molecular junctions can be achieved as the photoreaction transforms the wedge shaped SO-X to the coplanar conformation (MO-X). The nitro and amino substitutions can both increase ON/OFF ratio of MO-X/SO-X, and the MO-A/SO-A has the highest efficiency in comparison with other two groups. We can predict that the SO-X/MO-X can be a significant candidate for the molecular switch. According to what we have studied, different groups can be chosen as the highest-efficiency molecular switch under different voltages.

**Acknowledgements**

This work was supported by the National Undergraduate Scientific and Technological Innovation Project of China (Grant No. 201210422010).